\documentclass[12pt]{article}
\usepackage{a41}

\usepackage{palatino}
\usepackage{xcolor}
\usepackage{amsmath}
\usepackage{booktabs}
\usepackage{microtype}

\usepackage{cite}
\usepackage{amsmath}
\usepackage{amssymb}
\usepackage{color}

\usepackage[rflt]{floatflt}
\usepackage{float}
\usepackage{slashed}

\def\b0{\beta_0}

\usepackage{array}

\newcommand{\zb}{\overline{z}}

\usepackage[english]{babel}
\usepackage[latin1]{inputenc}
\usepackage[T1]{fontenc}
\usepackage{ae}

\usepackage{url}

\usepackage{amsmath, amsthm, amssymb}
\newtheorem{thm}{Theorem}[section]

\newtheorem{definition}[thm]{Definition}

\newcommand{\pvec}{{\rm\bf p}}
\newcommand{\rvec}{{\rm\bf r}}

\newcommand{\ep}{\varepsilon}

\usepackage{rotating}

\usepackage{graphicx}

\newcounter{mmacnt}
\def\restartmma{\setcounter{mmacnt}{0}}
\restartmma \catcode`|=\active
\def|#1|{\mathrm{#1}}
\catcode`|=12
\newenvironment{mma}{
 \par\smallskip
 \catcode`|=\active
 \parskip=0pt\parindent=0pt 
 \small
 \def\In##1\\{%
\def\linebreak{\hfill\break\null\qquad}%
\refstepcounter{mmacnt}
\hangindent=2.5em\hangafter=0
\leavevmode
\llap{\tiny\sffamily n[\arabic{mmacnt}]:=\kern.5em}%
\mathversion{bold}\footnotesize$\displaystyle##1$\normalsize
\mathversion{normal}\par
 }%
 \def\Print##1\\{%
\def\linebreak{\hfill\break}%
\hangindent=2.5em\hangafter=0
\leavevmode ##1\par}%
 \def\Out##1\\{%
\def\linebreak{$\hfill\break\null\hfill$}%
\kern\abovedisplayskip\par
\hangindent=2.5em\hangafter=0
\leavevmode
\llap{\tiny\sffamily Out[\arabic{mmacnt}]=\kern.5em}
\footnotesize$\displaystyle##1$\normalsize\hfill\null\par
\kern\belowdisplayskip
 }%
 \def\Warning##1##2\\{%
\def\linebreak{\hfill\break}%
\hangindent=2.5em\hangafter=0
\leavevmode
{\scriptsize##1 : ##2}\par}%
}{%
 \par\smallskip
}

\usepackage{color}

\newenvironment{fshaded}{%
\MakeFramed {\FrameRestore}
}%
{\endMakeFramed}

\def\b0{\beta_0}

\def\Gp0{{\Gamma^{'}_0}}
\def\Gp1{{\Gamma^{'}_1}}
\def\Gp2{{\Gamma^{'}_2}}

\allowdisplaybreaks[4]

\begin{document}
\setlength{\baselineskip}{0.515cm}

\sloppy
\thispagestyle{empty}
\begin{flushleft}
DESY 19--185
\hfill 
\\
DO--TH 19/21\\
SAGEX-19-25\\
\end{flushleft}

\mbox{}
\vspace*{\fill}
\begin{center}

{\Large\bf From Momentum Expansions to Post-Minkowskian} 

\vspace*{3mm}
{\Large\bf Hamiltonians by Computer Algebra Algorithms}

\vspace{3cm} \large
{\large J.~Bl\"umlein$^a$, A.~Maier$^a$, P.~Marquard$^a$, G.~Sch\"afer$^b$, and C. Schneider$^c$}

\vspace{1.cm}
\normalsize
{\it $^a$Deutsches Elektronen--Synchrotron, DESY,}\\
{\it   Platanenallee 6, D--15738 Zeuthen, Germany}

\vspace*{2mm}
{\it $^b$Theoretisch-Physikalisches Institut, Friedrich-Schiller-Universit\"at, \\
Max-Wien-Platz 1, D--07743 Jena, Germany}\\

\vspace*{2mm}
{\it $^c$Research Institute for Symbolic Computation (RISC),\\
                          Johannes Kepler University, Altenbergerstra\ss{}e 69,
                          A--4040 Linz, Austria}


\end{center}
\normalsize
\vspace{\fill}
\begin{abstract}
\noindent
The post-Newtonian and post-Minkowskian solutions for the motion of binary mass systems in gravity 
can be derived in terms of momentum expansions within effective field theory approaches. In the 
post-Minkowskian approach the expansion is performed in the ratio $G_N/r$, retaining all velocity terms 
completely, while in the post-Newtonian approach only those velocity terms are accounted for which are 
of the same order as the potential terms due to the virial theorem. We show that it is possible to obtain
the complete post-Minkowskian expressions completely algorithmically, under most general purely mathematical
conditions from a finite number of velocity terms and illustrate this up to the third post-Minkowskian 
order given in \cite{Bern:2019crd}.
\end{abstract}

\vspace*{\fill}
\noindent
\newpage

\section{Introduction}
\label{sec:1}

\vspace*{1mm}
\noindent
The use of a non-relativistic effective field theory 
\cite{EFT,Kol:2007bc,Gilmore:2008gq,Foffa:2011ub,Foffa:2019rdf,Foffa:2019yfl,Foffa:2019hrb,
Blumlein:2019zku}, provides one way to derive the equations of motion of a binary mass system within the post-Newtonian (PN) approach. 
Currently all corrections are known up to the 4th post-Newtonian order \cite{Kol:2007bc,Gilmore:2008gq,Foffa:2011ub,Foffa:2019rdf,Foffa:2019yfl} 
and first corrections due to the static potential at 5PN \cite{Foffa:2019hrb,Blumlein:2019zku}. Previously, the results up to the 
4th post-Newtonian 
order have already been derived using different methods, see Refs.~\cite{Damour:2014jta,Bernard:2017ktp} and references 
therein, and first results up to the 5th post-Newtonian order have been obtained in \cite{Bini:2019nra} recently.

The general principle is to expand in the ratio $G_N/r$, with $G_N$ Newton's constant and $r$
denoting the distance between the two point masses $m_1$ and $m_2$, and to retain all velocity corrections up to the
order implied by the virial theorem \cite{CLAUSIUS} $G_N m_1 m_2/r \sim  {\rm \bf v}_1^2 m_1 + {\rm \bf v}_2^2 m_2$. In the 
post-Minkowskian (PM) 
approach \cite{Bern:2019crd,PM,Cheung:2018wkq,Antonelli:2019ytb} the expansion is performed in $G_N/r$ retaining all 
velocity corrections 
as closed form expressions.
Recently calculations have been performed up to the third post-Minkowskian order, cf.~\cite{Bern:2019crd,Cheung:2018wkq}, in the 
center 
of momentum frame using isotropic coordinates $\pvec.\rvec = 0$. Here the Hamiltonian reads\footnote{It has been shown in 
\cite{Antonelli:2019ytb} that the results of \cite{Bern:2019crd} are in accordance with the energy- and angular momentum-orbital 
frequency relation, $E(\omega)$ and $L(\omega)$, obtained in the post-Newtonian approach up to $O((G_N M \omega)^2)$, with $M = m_1 
+m_2$. We have verified that the results in \cite{Bern:2019crd}, when compared to the post-Newtonian results in ADM coordinates 
in \cite{Damour:2014jta}, agree up to terms of $O(G_N^3 (\pvec^2)^2)$.}
\begin{eqnarray}
\label{eq:POT1}
H(\pvec,\rvec) &=& 
\sqrt{m_1^2 + \pvec^2} + \sqrt{m_2^2 + \pvec^2} + V(\pvec,\rvec),\\
\label{eq:POT1}
V(\pvec,\rvec) &=&  \sum_{k=1}^\infty V_k(\pvec) \frac{G_N^k}{|\rvec|^k},~~~~
V_k(\pvec) = \sum_{l=0}^\infty a_{k}(l) x^l,
\end{eqnarray}
where $x$ denotes an appropriate expansion variable, which will be defined below.

The question arises, whether these corrections can also be obtained using the effective field theory approach, in which one usually 
can only expand up to finite terms in the velocity. In this note we show that this is indeed possible algorithmically under the 
following three sufficient conditions.
\begin{enumerate}
\item There exist recurrences for the coefficients $a_k(l)$ of Eq.~(\ref{eq:POT1}) in $l$, up to a finite number of 
polynomial terms 
in $x$.
\item The recurrence or its associated differential equation factorizes at first order. 
\item The dependence of $V_k(x)$ on $\rho = m_1/m_2$ is rational.
\end{enumerate}
Here we will not use any special additional physical conditions, e.g. on expected structures of the solution, 
but follow a purely mathematical approach instead. 
We will use the method of guessing, see e.g. \cite{GSAGE}, to obtain the corresponding difference equation, which is then 
solved by applying difference field theory as implemented in the package {\tt Sigma} \cite{Schneider:2007a,Schneider:2013a}.
The final expressions are then obtained by performing one infinite sum and adjusting one polynomial by initial conditions.
In perturbative Quantum Chromodynamics (QCD) the method of guessing has been successfully applied to problems which are much more 
voluminous then the present ones, see Refs.~\cite{GUESS1,Blumlein:2017dxp}, and are based on up to $O(5000-8000)$ input values.

Concerning the integration, the master integrals in the zero-dimensional case are simpler to perform than for the momentum 
resummed
expressions. Moreover, all the master integrals in the case one expands in the momentum are already known up to five-loop order
from the post-Newtonian approach. The challenge for the present method lies in the expansion up to moderately large powers in the 
momentum.

In the effective field theory approach one usually starts out to work in harmonic coordinates. Here higher order time derivatives 
of the 
velocities occur, which may be eliminated by using the equation of motion to obtain a first order Lagrangian resp. Hamiltonian. 
This operation also induces a coordinate transformation \cite{Damour:1990jh}. One may then transform to coordinates which are 
connected to ADM coordinates \cite{Damour:2014jta}.The structure of the transformation matrices between the different systems 
can be fixed by confronting a corresponding ansatz with the results for observables being calculated in both frames up to the 
desired perturbative order. Here one well suited observable is the scattering angle \cite{DAM1}.

In the following we will demonstrate how the potentials of Eq. (2) can be completely recovered algorithmically from a finite number
of expansion coefficients $a_k(l)$. We will first consider the equal mass case $m_1 = m_2$, and then turn to the general 
case. In an appendix we discuss how some special sums occurring can be carried out. 
\section{The equal mass case}
\label{sec:2}

\vspace*{1mm}
\noindent
We first study the case $m_1 = m_2 \equiv m$, introduce the variable 
\begin{eqnarray}
x = \frac{\pvec^2}{m^2}
\end{eqnarray}
and normalize $V_k$ by a factor $1/m_1^{k+1}$ to obtain dimensionless quantities. We will keep this normalization
in the unequal mass case as well.

Within the effective 
field theory approach we sort the contributions keeping the terms of $O((G/r)^k)$ in the $k$th post-Minkowskian 
order and retain a finite number of terms in $\left.x^l\right|_{l=0}^M$, with $M$ a sufficiently large integer, 
which will be specified below. Next we seek for a recursion relation for the coefficients $a_{k}(l) \in \mathbb{Q}$ 
in 
$l$ for $k = 1,2,3$. 
We apply the method of guessing to a finite set of theses coefficients, which allows one to obtain this recurrence
of order {\sf o} and degree {\sf d},
\begin{eqnarray}
\sum_{n=0}^{\sf o} Q_n(l) f[l+n] = 0.
\end{eqnarray}
and to check its validity. Here $Q_n$ are polynomials in $l$ maximally of degree {\sf d}. Whenever the 
corresponding recurrences are first order 
factorizable in difference fields, they can be solved in terms of iterated sums defined over hypergeometric products 
by the package {\tt Sigma} \cite{Schneider:2007a,Schneider:2013a}. For this
{\sf o} initial values are needed, which are given by a subset of the expansion coefficients $a_{k}(l)$.
In this way we obtain a closed form for the expansion coefficients $a_{k}(l)$, possibly 
up to a finite number of expansion terms in $x$. The latter function is a polynomial.
By performing a Taylor series expansion of the result, one can fix these terms. Finally, we perform the infinite sum
analytically and obtain the closed form expressions for the potentials at the $k$th post-Minkowskian order.
Here the important point is, that the reconstruction is possible using a finite number of terms.

The simplest example is $V_1$. The momentum expansion yields the series
\begin{eqnarray}
V_1(x) &\simeq&
-1-7 x-x^2+x^3-x^4+x^5-x^6+x^7-x^8+x^9-x^{10}+x^{11}-x^{12}+x^{13}-x^{14}
\nonumber\\
& & +x^{15}
-x^{16}+x^{17}-x^{18}+x^{19}-x^{20}
+ O(x^{21}),
\\
S_1 &=& \{-1,-7,-1,1,-1,1,-1,1,-1,1,-1,
           1,-1,1,-1,1,-1,1,-1,1,-1\}.
\end{eqnarray}
In the effective field theory approach one would obtain the sequence $S_1$ by expanding in $x$ into a formal
Taylor series.

By guessing we obtain the recurrence
\begin{eqnarray}
f_1[n] + f_1[n+1] = 0,
\end{eqnarray}
with the solution
\begin{eqnarray}
\bar{a}_1(l) = (-1)^{l+1}.
\end{eqnarray}
Here the potential (\ref{eq:POT1}) has the representation
\begin{eqnarray}
\label{eq:POT2}
V_k(\pvec) = p_k(x) +
\sum_{l=0}^\infty \bar{a}_{k}(l) x^l,
\end{eqnarray}
where $p_k(x)$ is a polynomial. The coefficients of $p_k(x)$ are determined comparing the Taylor expansions
of (\ref{eq:POT2}) with those of (\ref{eq:POT1}). For this step we need 8 initial values. The corresponding solution reads then
\begin{eqnarray}
V_1 = p_1(x) - \frac{1}{1+x},
\end{eqnarray}
and $p_1(x)$ is given by
\begin{eqnarray}
p_1(x) = -8 x.
\end{eqnarray}
The same algorithm can now be applied at 2PM and 3PM.
In Table~\ref{TAB1} we list the respective numbers of minimally necessary input values to establish the corresponding
difference equations without any further assumption together with their orders and degrees. 
\begin{table}[H]
\centering
\renewcommand{\arraystretch}{1.5}
\begin{tabular}{|c|r|r|r|}
\hline
\multicolumn{1}{|c}{} & 
\multicolumn{1}{|c}{order} & 
\multicolumn{1}{|c}{degree} & 
\multicolumn{1}{|c|}{\# input values}  \\
\hline
1PM & 1  &  0  &  8   \\
2PM & 2  &  5  & 24   \\
3PM & 3  & 15  & 54   \\
\hline
\end{tabular}
\renewcommand{\arraystretch}{1}
\caption[]{\sf \small Characteristics of the recurrences and number of input values for the different post-Minkowskian 
orders in the equal mass case.}
\label{TAB1}
\end{table}
The exact value of the minimal number of input parameters needed at low numbers, as is the case here, is
determined experimentally. Usually one works with a larger number of expansion coefficients. Here we wanted 
to display the minimal value needed, since the expansions in $x$ in the effective field theory approach 
is not yet possible fully algorithmically. 

The recurrences for $V_2$ and $V_3$ read
\begin{eqnarray}
&& (10713-278 n-3688 n^2-400 n^3+16 n^4-96 n^5) f_2[n]
\nonumber\\ &&
+(4461-10188 n-8648 n^2-672 n^3
+80 n^4-192 n^5) f_2[n+1]
\nonumber\\ &&
+(-6252-9910 n-4960 n^2-272 n^3+64 n^4-96 n^5) f_2[n+2]  = 0
\\
&&
  Q_1~f_3[n] 
+ Q_2~f_3[n+1]
+ Q_3~f_3[n+2] 
+ Q_4~f_3[n+3]  = 0,
\end{eqnarray}
with
\begin{eqnarray}
Q_1 &=&
    126903309120 + 327090111984 n +
    199501827192 n^2 - 15839063268 n^3 
\nonumber\\ &&
+ 125598633964 n^4 +
    319201064194 n^5 + 244500413870 n^6 + 74947793534 n^7 
\nonumber\\ &&
- 2304037362 n^8 -
    7916007828 n^9 - 1912314952 n^{10} - 69778816 n^{11} 
\nonumber\\ &&
+ 36357088 n^{12} +
    6925120 n^{13} + 938880 n^{14} + 84480 n^{15},
\\
Q_2 &=&
   120213548280 + 370215834660 n + 215909250030 n^2 - 222044191596 n^3 
\nonumber\\ &&
-
    29242273581 n^4 + 508977450525 n^5 + 530659013385 n^6 +
    196490815287 n^7 
\nonumber\\ &&
+ 1030457202 n^8 - 20244382200 n^9 - 5290755840 n^{10} -
    189159456 n^{11} 
\nonumber\\ &&
+ 115021824 n^{12} + 22664640 n^{13} + 2985600 n^{14} +
    253440 n^{15}, 
\\
Q_3 &=&
   -101452470840 - 39208126464 n + 158218785864 n^2 - 262227440529 n^3 
\nonumber\\ &&
-
    511020293886 n^4 + 6335298708 n^5 + 336175588032 n^6 + 179465093733 n^7 
\nonumber\\ &&
+
    10546876182 n^8 - 17246382984 n^9 - 4999683456 n^{10} - 174901344*n^{11} 
\nonumber\\ && 
+
    123193344 n^{12} + 24786240 n^{13} + 3154560 n^{14} + 253440 n^{15}, 
\\
Q_4 &=& 185299500960 + 326857962960 n + 377719564176 n^2 + 3846793164 n^3 
\nonumber\\ &&
-
    343897751366 n^4 - 172310662748 n^5 + 55100830352 n^6 + 58117627580 n^7 
\nonumber\\ &&
+
    6784432878 n^8 - 5022804012 n^9 - 1624176328 n^{10} - 53736544 n^{11} 
\nonumber\\ &&
+
    44718688 n^{12} + 9046720 n^{13} + 1107840 n^{14} + 84480 n^{15}.
\end{eqnarray}
We find for $\bar{a}_2(l)$ and $\bar{a}_3(l)$
\begin{eqnarray}
\bar{a}_2(l) &=& 
6 (-1)^{l+1}
-\frac{\big(
        87+96 l+168 l^2+128 l^3-16 l^4\big)} {(2 l-3) (2 l-1)} \left(\frac{-1}{4}\right)^{l+1} \frac{(2l)!}{(l!)^2},
\\
\bar{a}_3(l) &=& 
\frac{1}{6} \big(
        -114-94 l-15 l^2+l^3\big) (-1)^l
+\frac{\big(
        12+22 l+l^2+14 l^3+11 l^4\big) 
}{(l-1) l (1+2 l) (3+2 l)} 
\frac{(-1)^l 2^{3+2 l} (l!)^2}{(2 l)!}
\nonumber\\ &&
+\frac{3 (5+2 l)\big(
        -283-470 l-312 l^2-40 l^3+16 l^4\big) 
}{(1+l) (2+l) (-1+2 l)} 
\frac{(-1)^l 2^{-2-2 l} (2 l)!}{(l!)^2}.
\end{eqnarray}
Note that the expression for $\bar{a}_3$ is valid for $l > 1$ only. The remaining sum for $\left.V_i(x)\right|_{i=1}^3$
can be simply performed using {\tt Mathematica} and more specialized software, which may be necessary for higher
post-Minkowskian orders, is not yet needed.

The lower coefficients can be determined using the corresponding input values. The polynomials $p_{2(3)}(x)$ are
\begin{eqnarray}
p_2(x) &=& - 30 x,
\\
p_3(x) &=& -\frac{5}{2}-\frac{1027 x}{8}-\frac{43917 x^2}{80}.
\end{eqnarray}
By using (\ref{eq:POT1}) we finally reconstruct the functions given in \cite{Bern:2019crd} Eqs.~(10.10) 
in the equal mass case:\footnote{The logarithm in (\ref{eq:V3}) can be transformed into a arcsinh-function.}
\begin{eqnarray}
V_1 &=&  -8 x - \frac{1}{1 + x},
\\
V_2 &=&  
-\frac{6 \big(
        1+5 x+5 x^2\big)}{1+x}
+\frac{1}{4} \frac{\big(
        1+8 x+8 x^2
\big)
\big(29+72 x+40 x^2\big)}{(1+x)^{5/2}},
\\
V_3 &=&
-\frac{3424 x^2}{3}
+\frac{3 \big(
        -1+\sqrt{1
        +x
        }\big)}{x}
+\frac{80}{3} x \big(
        -25+27 \sqrt{1
        +x
        }\big)
\nonumber\\ &&
+\frac{-1
+8 (1+x)
-23 (1+x)^3
-68 (1+x)^4
+3 \big(
        34+22 x-55 x^2-40 x^3\big) (1+x)^{3/2}
}{(1+x)^4}
\nonumber\\ &&
+ \frac{2 \log \big(
        \sqrt{x}
        +\sqrt{1+x}
\big)
\big[-11
        +16 x (1+x) (-1
        +4 x (1+x)
        )
\big]}{\sqrt{x} (1+x)^{3/2}}.
\label{eq:V3}
\end{eqnarray}
\section{The general case}
\label{sec:3}

\vspace*{1mm}
\noindent
While the determination of multi-variate recurrence relations is possible in certain cases, their solution
is much more difficult than the one in the uni-variate case and the theory behind is less known.
Therefore, we propose a different way to solve the general case. In performing the momentum series
in isotropic coordinates we can keep the masses different without any effort. This allows to study
the reconstruction similar to the one in Section~\ref{sec:2} after fixing the ratio $\rho = m_1/m_2$ to a rational 
number or an integer. Except particular degenerate cases one will always find the same functional structure in a 
chosen kinematical variable $\xi$. The choice of primes for $\rho$ serves this purpose.

Starting with the asymmetric kinematic variable $\xi = \pvec^2/m_1^2$ the kinematic square roots 
$\sqrt{m_i^2 + \pvec^2},~i = 1,2$ remain. They cause a problem in the reconstruction, since the expansion coefficients 
for the function
\begin{eqnarray}
\sqrt{1+x} \sqrt{1+\rho^2 x} = \sum_{k=0}^\infty b(k) x^k
\end{eqnarray}
are
\begin{eqnarray}
b(k) = \sum_{l=0}^k \binom{\tfrac{1}{2}}{l} \binom{\tfrac{1}{2}}{k-l} \rho^{2l} = \binom{\tfrac{1}{2}}{k} {}_2F_1\left[
\renewcommand{\arraystretch}{1.3}
\begin{array}{c} -\tfrac{1}{2},~-k\\ \tfrac{3}{2} - k \end{array}
\renewcommand{\arraystretch}{1}
; \rho^2
\right].
\end{eqnarray}
$b(k)$ is not a hypergeometric term since $b(k+1)/b(k)$ is not a rational function (of fixed numerator and 
denominator degree). One may obtain a recurrence for $b(n)$ but it is not solvable within our available difference field
algorithms. 

We rather choose a more physical variable appropriate for the two--mass case\footnote{One often does this in massive 
calculations. Landau variables are early examples for this \cite{LANDAU}. Of course, after this change of variable 
also related integrals can then be performed in simpler function spaces \cite{Blumlein:2018cms}.},
\begin{eqnarray}
\label{eq:y1}
z(x) &=& \frac{1 + \rho^2}{4 \rho} + \frac{\rho x}{2} + \frac{1}{2} \sqrt{1 + x} \sqrt{1 + \rho^2 x},~~~x = \frac{\pvec^2}{m_1^2}
\\
  &=& \frac{(E_1 + E_2)^2}{4 m_1 m_2},
\end{eqnarray}
with $E_i = \sqrt{\pvec^2 + m_i^2}$,
cf. also \cite{SCHELE}, which has been also used in \cite{DY} recently. The inversion of (\ref{eq:y1}) reads
\begin{eqnarray}
x \equiv \frac{[(1 - \rho)^2 - 4 \rho z][(1 + \rho)^2 - 4 \rho z]}{16 \rho^3 z} = \frac{(E^2-M_+^2)(E^2-M_-^2)}{4 E^2 m_1^2},
\end{eqnarray}
where $E = E_1 + E_2$ and $M_\pm = m_1 \pm m_2$.

It is also useful to change to the variable $\overline{z}$, given by
\begin{eqnarray}
\bar{z} = z - \frac{(m_1 + m_2)^2}{4 m_1 m_2},
\end{eqnarray}
which vanishes for $\pvec^2 \rightarrow 0$ like the expansion variable in the equal mass case.
Note also that in these variables one obtains the contributions to the potential in a very compact form.

Now, the further dependence on $\rho$ is at most a rational function given by integer coefficients.
\begin{table}[H]
\centering
\renewcommand{\arraystretch}{1.5}
\begin{tabular}{|c|r|r|r|}
\hline
\multicolumn{1}{|c}{} & 
\multicolumn{1}{|c}{order} & 
\multicolumn{1}{|c}{degree} & 
\multicolumn{1}{|c|}{\# input values}  \\
\hline
1PM & 2 &  0 &   8  \\
2PM & 4 & 12 &  45  \\
3PM & 9 & 26 & 120  \\
\hline
\end{tabular}
\renewcommand{\arraystretch}{1}
\caption[]{\sf \small Characteristics of the recurrences and number of input values for the different post-Minkowskian 
orders in the general case.}
\label{TAB2}
\end{table}

Let us consider the cases $m_1 =3  m_2$ and $m_1 =5  m_2$  for  $V_1$  as examples. 
We obtain the recurrences
\begin{eqnarray}
9 f[n] - 4 f[2 + n] = 0,~~~~25 f[n] - 36 f[2 + n] = 0,
\end{eqnarray}
from $8$ initial values, which are one order higher than in the equal mass case. This is potentially expected, 
since certain structural simplifications w.r.t. the variable $x$ can occur in the equal mass case. Still we 
always seek the lowest order recurrences. By similar steps as performed in Section~\ref{sec:2} we obtain
\begin{eqnarray}
V_1\left(z; \rho = 3\right) = 
-\frac{128 - 480 z + 369 z^2}{27 z (4-9z^2)},~~~
V_1\left(x; \rho = 5\right) = 
-\frac{10368 - 18720 z + 7825 z^2}{125 z (36 - 25 z^2)},
\nonumber\\
\end{eqnarray}
with the polynomial contributions
\begin{eqnarray}
p_1^{(2)}(z;\rho = 3) = \frac{41}{27} \frac{1}{z},~~~~
p_1^{(2)}(z;\rho = 5) = \frac{313}{125} \frac{1}{z}~.
\end{eqnarray}
For sufficiently asymmetric choices of rational ratios of the masses this structure remains and is given by
\begin{eqnarray}
V_1\left(z; \frac{m_1}{m_2} = \rho\right) = 
\frac{c_1(\rho) + c_2(\rho) z + c_3(\rho) z^2}{\rho^3 z \big[
        \big(
                1-\rho^2\big)^2
        -16 \rho^2 z^2
\big]}. 
\end{eqnarray}
Here the dependence on $\rho$ in the denominator is easily visible. 
The functions $c_i(\rho),~i=1,2,3$ are polynomials up to degree $d=8$. One obtains
\begin{eqnarray}
V_1(z;\rho)
&=& 
\frac{-(1 - \rho^2)^4  + 8 (1 - \rho^2)^2 \rho  (1 + \rho^2) z - 
 16 \rho^2 (1 + \rho^4) z^2}
{2 \rho^3 z [(1 - \rho^2)^2 - 16 \rho^2 z^2]}.
\end{eqnarray}
To determine the polynomial coefficients for $\rho$ in the numerators, one removes the denominator and solves 
the corresponding system of linear equations, which one extends as long as the coefficient matrix is not degenerate. 
Using the variable $\overline{z}$ instead of $z$ will give a similar result in the case of $V_1$.
Finally, one transforms back from $z$ to $x$, cf.~(\ref{eq:y1}).

In a similar way one proceeds for $V_2$ and $V_3$. Here we display the results for $\rho =3$ only. The same structures with 
different values of the coefficients are obtained choosing other odd prime ratios $\rho$.
Here and in the following we use the variable $\overline{z}$. 
One obtains 
\begin{eqnarray}
V_2(\overline{z}; \rho = 3) &=& 
-\frac{4 P_5}{(2+\overline{z})^2 (4+3 \overline{z})}
+ \frac{P_6 P_7 }{12 (2+\overline{z})^6 (4+3 \overline{z})^{5/2}},
\end{eqnarray}
with
\begin{eqnarray}
P_5 &=& 5 \overline{z}^4+25 \overline{z}^3+49 \overline{z}^2+44 \overline{z}+16,
\\
P_6 &=&8 \overline{z}^4+40 \overline{z}^3+73 \overline{z}^2+56 \overline{z}+16,
\\
P_7 &=& 360 \overline{z}^6+3720 \overline{z}^5+15629 \overline{z}^4+34588 \overline{z}^3+42804 
\overline{z}^2+28192 \overline{z}+7744.
\end{eqnarray}
The polynomial not obtained by guessing reads
\begin{eqnarray}
p_2^{(2)}(\overline{z};\rho = 3) = \frac{20}{9} - \frac{20}{3} \overline{z}.
\end{eqnarray}
The numerator polynomials of type $P_5$ have degree 5  and those of $P_6 \cdot P_7$ have $4 \times 8$ in $\rho$. 

Likewise, one proceeds in the case of $V_3$. 
\begin{eqnarray}
V_3(\overline{z}; \rho = 3) &=& 
-\frac{(4+3 \zb) P_{10}}{27 \zb (1+\zb) (2+\zb)^5 (2+3 \zb)^5}
+\frac{2 P_9~(4+3 \zb)^{5/2}}{9 \zb (1+\zb) (2+\zb)^3 (2+3 \zb)^3} 
\nonumber\\ &&
+\frac{2 (4+3 \zb) P_8}{9 (2+\zb) (2+3 \zb)} \frac{{\rm arcsinh}(\sqrt{\zb})}{\sqrt{\zb (1+\zb)}}, 
\end{eqnarray}
with the polynomials 
\begin{eqnarray}
P_8 &=& 64 \zb^4+128 \zb^3+48 \zb^2-16 \zb-11
\\
P_9 &=&720 \zb^7+4120 \zb^6+8718 \zb^5+9055\zb^4+4947 \zb^3+1363 \zb^2+156 \zb+4
\\
P_{10} &=& 277344 \zb^{13}+3690576 \zb^{12}+21772044 \zb^{11}+75259086 \zb^{10}
\nonumber\\ &&
+169789579 \zb^9+263423972 \zb^8+288178779 \zb^7+223910983 \zb^6+122572992 \zb^5
\nonumber\\ &&
+46092412 \zb^4+11326448 \zb^3+1657920 \zb^2+120832 \zb+3072.
\end{eqnarray}
The associated polynomial reads
\begin{eqnarray}
p_3^{(2)}(\overline{z};\rho = 3) = \frac{995}{216}-\frac{280333 \zb}{15552}-\frac{1526641 \zb^2}{23040}.
\end{eqnarray}
Some infinite sums are most economically calculated using relations of 
the type given in the appendix. Others are recognized as special $_pF_q$-functions, which finally reduce to 
elementary functions, cf.~\cite{PBM2}. The numerator polynomials in $\rho$ have a degree up to $d = 2$ for the arcsinh-term,
$d = 3$ for the term proportional to $(\tfrac{1}{4}(\rho+1)^2+\rho \zb)^{5/2}$, and
$d  = 12$ for the rational term. Finally, one substitutes back from $\overline{z}$ to $x$.

We mention that in \cite{Bini:2019nra,Antonelli:2019ytb,Damour:2015isa}, new special constants contribute which 
did not yet emerge in the above results. Terms resulting from the local-in-time dynamics introduce the new 
special constant $\pi^2$ from 3PN on which will show up in the post-Minkowskian series from 4PM on. 
The non-local-in-time dynamics, starting at 4PN, will introduce the new special constants $\ln(2), \ln(3), \ln(5)$  etc., together 
with the Euler-Mascheroni constant\footnote{Calculating loop-integrals in $D$ dimensions in momentum space, $\gamma_E$, only appears 
in the spherical factor $S_\ep$, which is set to unity at the end of the calculation working in the $\overline{\sf MS}$ scheme.
This seems not to be the case in gravity. Here the tail terms result in genuine $\gamma_E$ contributions.}, 
$\gamma_E$, the former of which are well-known to belong to the cyclotomic extension 
\cite{Ablinger:2011te} 
of the multiple zeta values \cite{Blumlein:2009cf}. Especially 
they are related to the (linear combinations) of the digamma function $\psi(k/l),~k,l \in \mathbb{N} \backslash \{0\}$ 
or its derivatives. Constants of this type and their associated one--dimensional functions emerge in many massive 
higher order calculations in QCD, cf.~\cite{Ablinger:2014bra,CYCL1}.
One should note that observables in the post-Newtonian approach, as e.g. $E(\omega)$, the energy-rotation 
frequency relation at the last stable orbit, contain also $\pi^2$ terms 
from 3PN onward, cf.~\cite{Damour:2014jta}. At 3PN they cannot come from the tail terms.
Additionally, from the non-local-in-time dynamics, the power expansion $(G_N/r)^n$ breaks
down and also receives contributions of $O[(G_N/r)^n{\rm ln}(G_N/r)]$.
\section{Conclusions}
\label{sec:4}

\vspace*{1mm}
\noindent
We have shown that one may determine the expansion coefficients of the potential in the post-Minkowskian approach
from the velocity expansion in the effective field theory approach by finite terms, using very general mathematical
algorithms. Here the approach works without any special further assumptions and even allows automation.
In the two--mass case the choice of appropriate symmetric variables is important. The recurrences determining 
the solutions for the potential up to the 3PM  are relatively compact compared to the ones needed in characteristic examples 
known from massless and massive calculations at three-loop order in QCD \cite{GUESS1,Blumlein:2017dxp}. Correspondingly, the 
solutions can 
be found by  comparatively low numbers of initial values and the reconstruction of the solutions proceeds very fast and does 
not require large computational resources. The method of guessing \cite{GSAGE}, implemented in {\tt Sage} \cite{SAGE}, 
delivers the recurrences for the expansion coefficients for the potentials $V_k$, which can be solved using the package 
{\tt Sigma} \cite{Schneider:2007a,Schneider:2013a}. The final reconstruction requires to perform one more infinite sum and the 
adjustment of one polynomial. This method can also be applied to the calculation of individual amplitudes, if their
momentum expansion can be performed. Here one can refer to the respective differential equations in the expansion parameter
and apply the method of large moments introduced in Ref.~\cite{Blumlein:2017dxp}.

\appendix
\section{Appendix}
\label{sec:A}

\vspace*{1mm}
\noindent
Let us consider the calculation of the following typical example sums which appear in the reconstruction of $V_3$ in the 
unequal mass case
\begin{eqnarray}
\sigma_1(x) & = & \sum_{n=1}^\infty n^2 \frac{(-x)^n}{2^{3+n}} \sum_{k=1}^n \frac{3^k \binom{2 k}{k}}{(2 k-1) 2^{3k}}, \\
\sigma_2(x) & = & \sum_{n=1}^\infty \left(\frac{-x}{2}\right)^n
\sum_{k=1}^n \frac{2^{3 k} }{\binom{2 k}{k} k}.
\end{eqnarray}
One uses the identity
\begin{eqnarray}
\sum_{n=1}^\infty x^n \sum_{l=1}^n f(l) = \frac{1}{1-x} \sum_{n=1}^\infty f(n) x^n.
\end{eqnarray}
Relations of this kind were considered in \cite{Fleischer:1998nb,Ablinger:2014bra} before. Furthermore,
\begin{eqnarray}
\sum_{n=1}^\infty n x^n f(n) = x \frac{d}{dx} \sum_{n=1}^\infty x^n f(n) 
\end{eqnarray}
holds, which is used to absorb the powers of $n$.

One finally obtains
\begin{eqnarray}
\sigma_1(x) & = & \frac{x
\big\{
        32
        +x \big[
                92
                + 3 x  \big(
                        12
                        -3 x
                        +8 \sqrt{4+3 x}
                \big)
                -16 \sqrt{4+3 x}
        \big]
        -64 \sqrt{4+3 x}
\big\}} 
{32 (2+x)^3 (4+3 x)^{3/2}},\\
\sigma_2(x) & = & -\frac{4x}{\sqrt{1+x}(2+x)} 
\frac{{\rm arcsinh}[\sqrt{x}]}{\sqrt{x}}.
\end{eqnarray}

\vspace*{5mm}
\noindent
{\bf Acknowledgment.} We thank Th.~Damour, K.~Sch\"onwald and J.~Steinhoff for discussions and Z.~Bern 
for a communication on Ref.~\cite{Bern:2019crd}. This work has been funded in part by the Austrian Science 
Fund (FWF) grant SFB F50 (F5009-N15), by EU TMR network SAGEX agreement No. 764850 (Marie Sk\l{}odowska-Curie) 
and COST action CA16201: Unraveling new physics at the LHC through the precision frontier. 

{\small

\end{document}